\documentclass[journal]{IEEEtran}
\usepackage{graphicx}
\usepackage{booktabs}
\usepackage{tabularx}
\usepackage{amsmath}
\usepackage{amssymb}

\ifCLASSINFOpdf
\else
\fi
\begin{document}
%
\title{Development of the Front-End Electronics\\     for PandaX-III Prototype TPC}
%
%
%

\author{Danyang~Zhu,
        Shubin~Liu,
        Changqing~Feng,
        Cheng~Li,
        Haolei~Chen
\thanks{This work is supported by the Special Program for Key Basic Research of Ministry of Science and Technology, China (Grant No. 2016YFA0400300), and the CAS Center for Excellence in Particle Physics (CCEPP).}
\thanks{D, Zhu, S. Liu, C. Feng, C. Li and H. Chen are from
State Key Laboratory of Particle Detection and Electronics, University of Science and Technology of China, Hefei 230026, China.}
\thanks{The authors are from Department of Modern Physics, University of Science and Technology of China, Heifei 230026, China.}
\thanks{Corresponding author: S. Liu, email: liushb@ustc.edu.cn.}}

%
%

\markboth{}%
{Zhu \MakeLowercase{\textit{et al.}}: Development of the Front-End Electronics for PandaX-III Prototype TPC}
%



\maketitle

\begin{abstract}
The Particle And Astrophysical Xenon Experiment III (PandaX-III) is an experiment to search for the Neutrinoless Double Beta Decay (NLDBD) using 200 kg radio-pure high-pressure gaseous xenon TPC with Micromegas detectors at both ends and cathode in the middle. A small-scale TPC equipped with 7 Microbulk Micromegas detectors is developed as the prototype detector. 128 channels are readout following an X-Y design with strips of 3 mm pitch (64 channels each direction) from one Micromegas module. Highly integrated front-end electronics composed of 4 front-end cards with 1024 readout channels is designed to readout the charge of Micromegas anode signals, digitize the waveform after shaping and send compressed data to the DCM board. The cornerstone of the front-end electronics is a 64-channel application specific integrated circuit which is based on a switched capacitor array. The integral nonlinearity of the front-end electronics is less than 1\% with 1 pC. The noise (RMS) of each readout channel is less than 0.9 fC with 1 $\mu$s peaking time and 1 pC range. Using the radioactive sources $^{56}$Fe and $^{137}$Cs, joint-tests of front-end electronics with the prototype TPC were carried out and the hit map of 7 Micromegas has been reconstructed. 
\end{abstract}


%
\IEEEpeerreviewmaketitle

\section{Introduction}
%
%
%
%
\IEEEPARstart{S}{earching} for the Neutrinoless Double Beta Decay
(NLDBD) is considered as a reliable approach to explore the nature of neutrinos. The PandaX-III experiment is the first large scale project aimed to search for the NLDBD of Xe-136 in a 200 kg high-pressure gas TPC which is filled with the mixture of $^{136}$Xe and TMA (trimethylamine). The experiment will be located in China Jin Ping Underground Laboratory (CJPL).

\begin{figure}
\includegraphics[width=3.5in,clip,keepaspectratio]{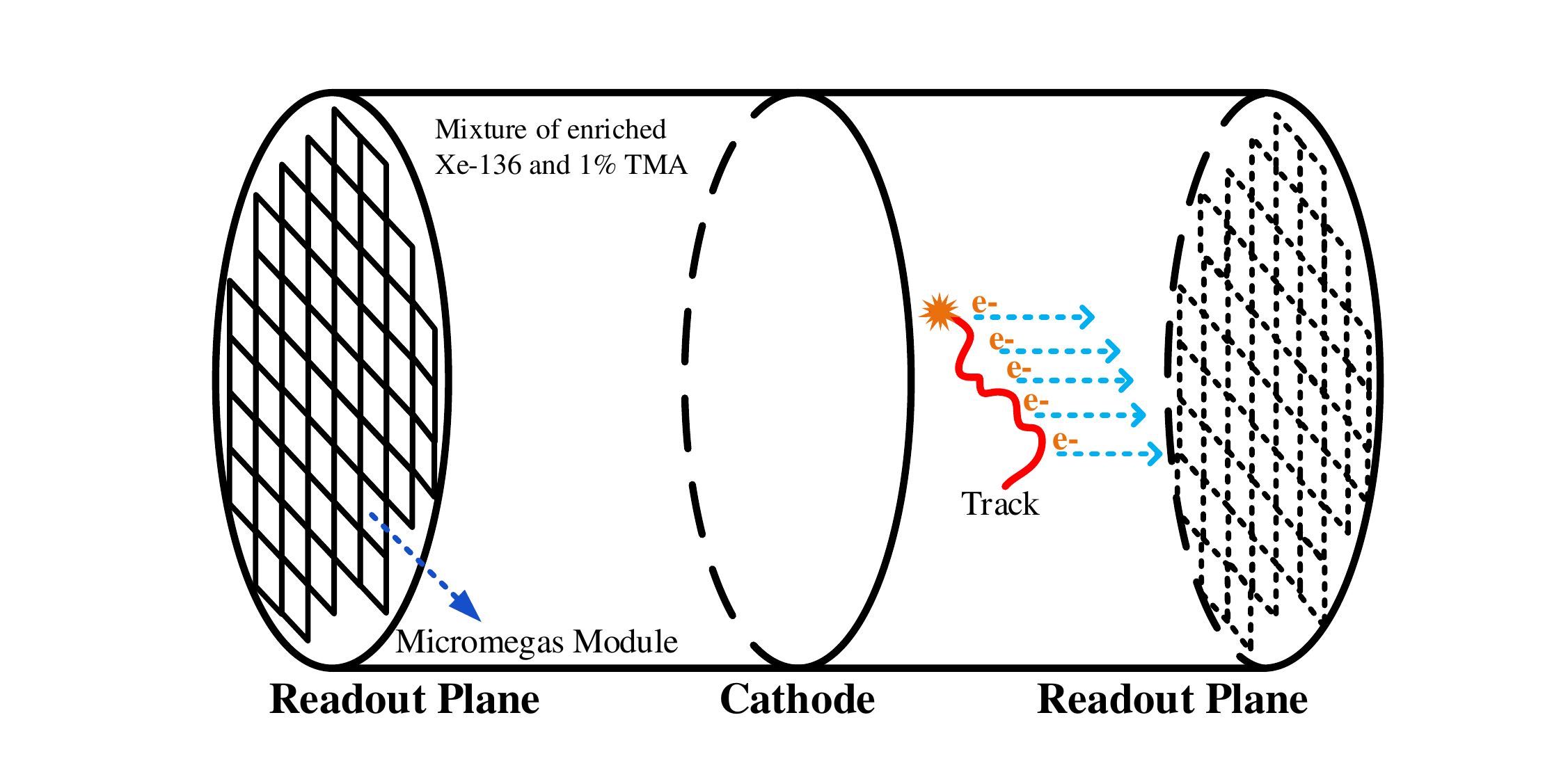}
\caption{Schematic of the PandaX-III TPC.}
\label{Fig. 1}
\end{figure}

The high pressure gas TPC, which is the central component of the PandaX-III detector, features a symmetric design with cathode in the middle and charge readout planes at the two ends, as shown in Fig.~\ref{Fig. 1} [1, 2, 3]. Tests at 10 bar of xenon + TMA mixture have shown energy resolution at the neutrino-less double beta decay Q-value down to 3\% FWHM [4]. And in the first phase of PandaX-III experiment, 20 cm $\times$ 20 cm Microbulk Micromegas detectors [5] are applied at both ends of the TPC to get better energy resolution. 

The prototype is a small-scale TPC equipped with 7 Microbulk Micromegas detectors.The prototype TPC is designed to measure event energy, track and other various features of PandaX-III detectors in Shanghai Jiaotong University (SJTU). To acquire anode signals of the Prototype TPC, an electronics system has been developed, in which the front-end electronics based on AGET chips is used to readout signals from the TPC Micromegas detectors.

\section{Readout Electronics System}
\subsection{Readout Requirements}
The prototype TPC is cylindrical, with single-side drift chamber, anode at the top and cathode at the bottom. The active volume is 66 cm in diameter and 78 cm tall, and 16 kg of xenon is contained within it at 10 bar. The size of each Microbulk Micromegas detector is 20 cm $\times$ 20 cm. 128 channels are readout following an X-Y design with strips of 3 mm pitch (64 channels each direction). Seven Microbulk Micromegas detectors are installed in the prototype TPC for a total of 896 channels to be read out.   

Except for the readout channels, there are some other requirements. Considering that 
the neutrino-less double beta decay Q-vaule of xenon-136 is about 2.5 MeV and the typical Micromegas amplification is approximately 1000 times, the input charge of all anode strips can be calculated as 10 pC [6]. According to simulation, when taking the extreme case that a 30 cm long electron track is being drifted towards the detector, the maximum charge in a strip is 1 pC. The Integral Non Linearity (INL) should be less than 3.2\%, the gain non-uniformity of all channels are less than 2\% and the Equivalent Noise Charge (ENC) is less than 6 fC.

Also taking the extreme case that a 30 cm long electron track is being drifted towards the detector, the maximum drift time is around 300 $\mu$s on the condition of the electrons only collected by one strip. However, by the Monte Cole simulation, it is found that the drift time of 95\% tracks in different events is less than 100 $\mu$s, and the charge pulse width of every anode strip signal in different events is less than 46 $\mu$s. 
\subsection{Basic Architecture}
\begin{figure}
\includegraphics[width=3.5in,clip,keepaspectratio]{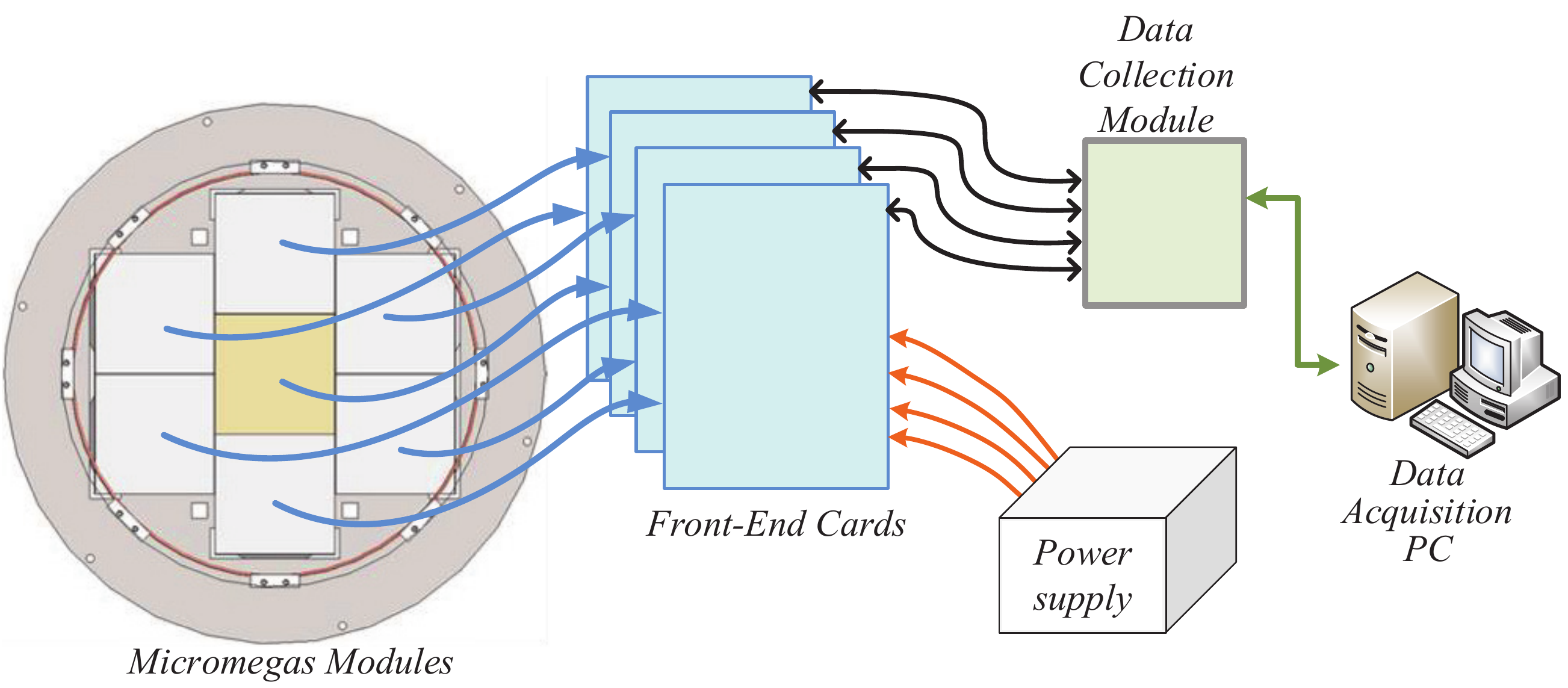}
\caption{Readout electronics system for PandaX-III prototype TPC.}
\label{Fig. 2}
\end{figure}

A readout electronics system has been put forward for the prototype TPC, which contains 4 Front-End Cards (FECs) and the Data Collection Module (DCM), as shown in Fig.~\ref{Fig. 2}. The FECs integrate and digitize input charges from anode strips of seven Micromegas detectors, and send the packaged data to the DCM. All the data from FECs can be gathered in the DCM and then transmitted to the data acquisition computer.

\section{Progress of Front-End Electronics}
\subsection{Design of the Front-End Electronics}
\subsubsection{Study of the AGET chip}
The charge pulse width of netrinoless double beta decay event can reach 46 $\mu$s, and it is difficult to directly sample such signals with long peaking time. However, using the pulse area to reconstruct the event energy instead of the peak amplitude is a feasible method, based on the simulation that the pulse areas of the signals with same charge are completely the same despite their different pulse width. After the investigation, the AGET chip is the most appropriate choice, because the readout time window of the AGET chip [7] can reach much more than 1 $\mu$s under the limitation of its peaking time.   

\begin{figure}
\includegraphics[width=3.5in,clip,keepaspectratio]{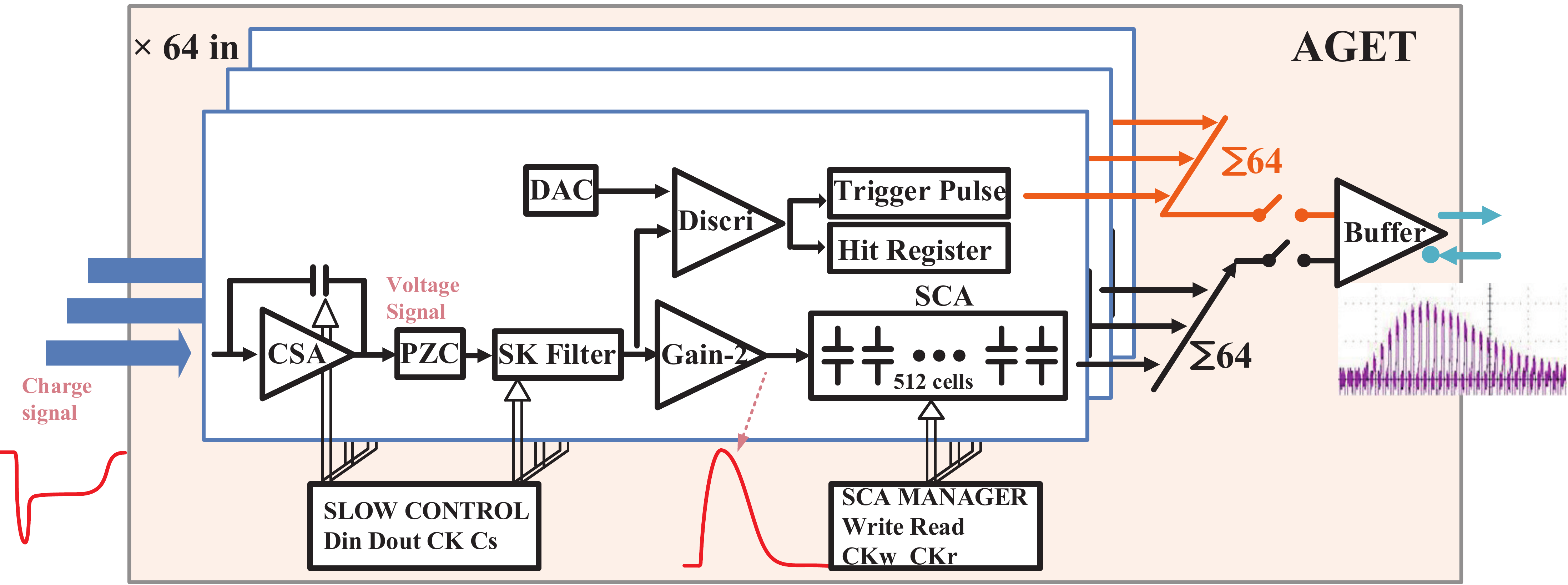}
\caption{Architecture view of the AGET cip, the figure is from [10].}
\label{Fig. 3}
\end{figure}

The AGET chip is based on the AFTER chip [8] with significant new features, and the latter one is used in the T2K experiment successfully [9]. The architecture of the AGET chip is shown in Fig.~\ref{Fig. 3} [10], it has 64 channels, and each channel integrates mainly: a charge sensitive amplifier (CSA), an analogue filter, a discriminator for trigger building and a 512-sample analog memory, which is based on a Switched Capacitor Array structure (SCA). The CSA has four available gains (120 fC, 240 fC, 1 pC, 10 pC). The analog filter is formed by Pole Zero Cancellation stage (PZC) followed by a 2-complex pole Sallen-Key low pass filter (SK filter) and the peaking time is selectable among 16 values in the range of 50 ns to 1 $\mu$s. Selecting internal polarization resistors allows operation with positive or negative input signals [11].  

The sampling frequency of the AGET chip can be set from 1 MHz to 100 MHz, and the readout is done at 25MHz. As the experimental requirement of input charge signal width is 46 $\mu$s and each channel has 512 evenly spaced points , the sampling rate of the AGET chip can be set to 5 MHz to reach the target. Although the maximum charge in the experiment is about 10 pC, the gain range can be set to 1 pC during the test. The reason for this is that the output of AGET is a wide pulse, whose amplitude is much smaller than that of the standard pulse. Besides, the total charge induced signals is shared amongst neighbouring readout strips.

\subsubsection{Design of the FEC}
\begin{figure}
\includegraphics[width=3.5in,clip,keepaspectratio]{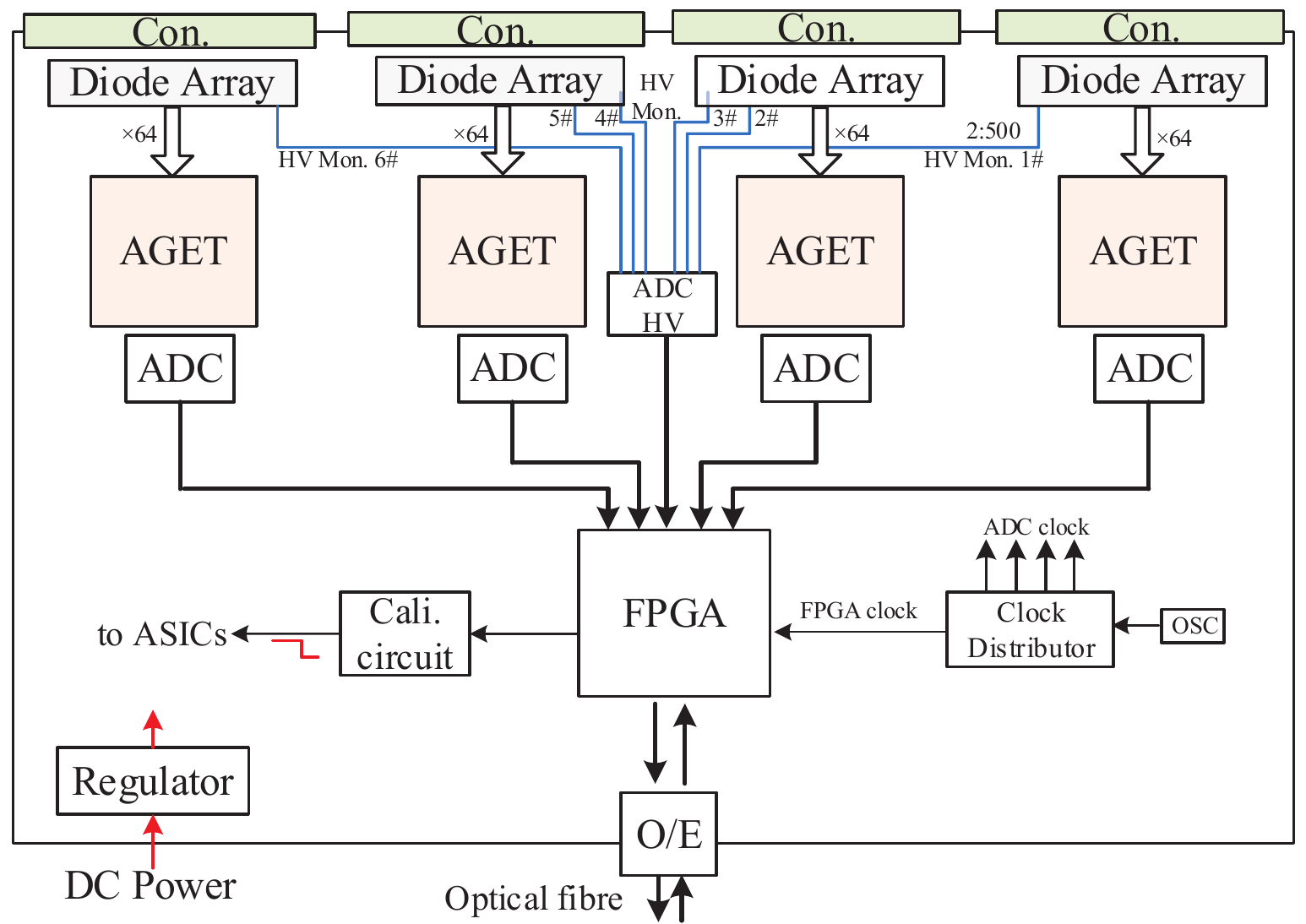}
\caption{Schematic diagram of the FEC, the figure is from [1].}
\label{Fig. 4}
\end{figure}

The Fig.~\ref{Fig. 4} shows that the FEC uses 4 AGET chips with 256 input channels to readout Micromegas anode signals via 4 dual row right angled 80-point 1.27 mm pitch surface mount connectors (Erni 154744) under the current protection of Electro-Static Discharge (ESD) protection circuits. The output of the AGET chips is digitized by four single-channel 12-bit ADCs (AD9235 from Analog Devices) clocked at 25 MHz, which are selected for the convenience of place and route [12]. The sampling clocks of the AGET chips and ADCs are provided by a multioutput clock distribution chip (AD9522 from Analog Devices) with subpicosecond jitter performance [13]. The synchronous pin of AD9522 can synchronize the sampling clocks of four AGET chips within 10 ns. 

\begin{figure}
\includegraphics[width=3.5in,clip,keepaspectratio]{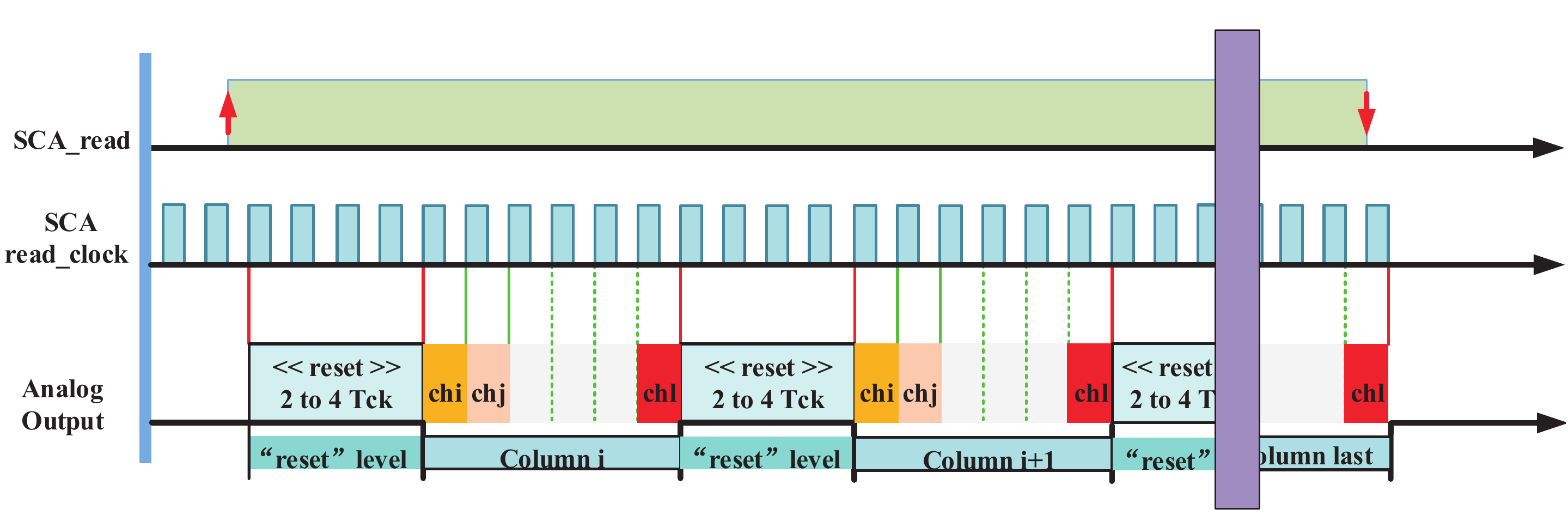}
\caption{Chronogram of the SCA read phase, the figure is from [10].}
\label{Fig. 5}
\end{figure}

\begin{figure}
\includegraphics[width=3.5in,clip,keepaspectratio]{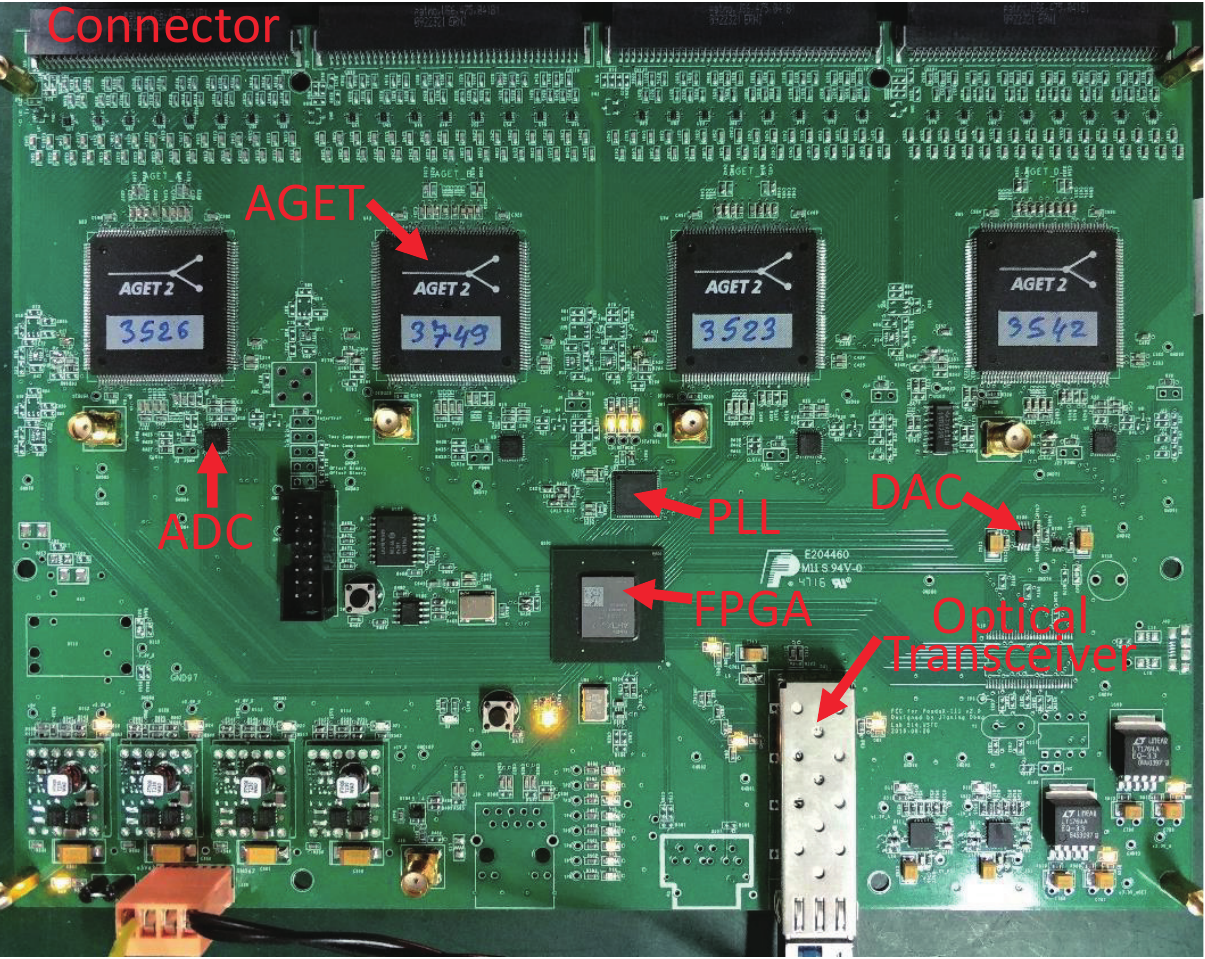}
\caption{Physical diagram of the FEC.}
\label{Fig. 6}
\end{figure}

The FPGA (Artix-7 from Xilinx) [14] aggregates analog data from 4 ADCs. The data sequence of 64 channels in the AGET chip is column by column, and each column contains one sampling point of 64 channels, as shown in Fig.~\ref{Fig. 5}. The FPGA has the ability to rearrange the data sequence with channel by channel to compress data in case for need, and store 4.2 Mbit data of two events in RAM. Supposing that the event rate is 10 Hz, the data rate of each FEC is 21 Mbps. Optical fibres are selected for the data transmission, command configuration and trigger signals distribution between the FEC and the DCM. In addition, the FEC includes an on-board pulser for channel linear calibration. A photograph of the FEC is shown in Fig.~\ref{Fig. 6}.    

\subsection{Performance Tests}
\begin{figure}
\includegraphics[width=3.5in,clip,keepaspectratio]{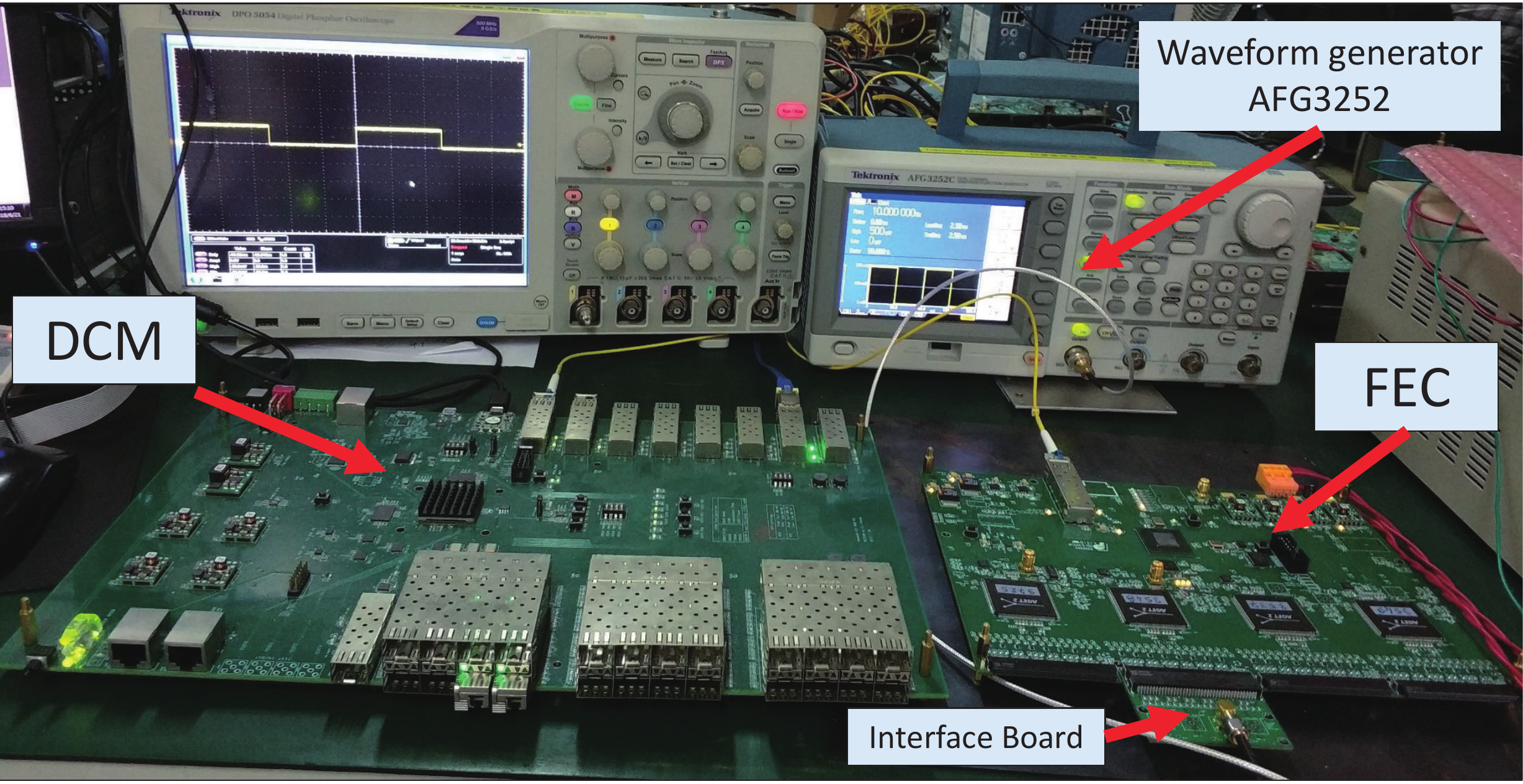}
\caption{A photograph of the test platform.}
\label{Fig. 7}
\end{figure}

\begin{figure}
\includegraphics[width=3.5in,clip,keepaspectratio]{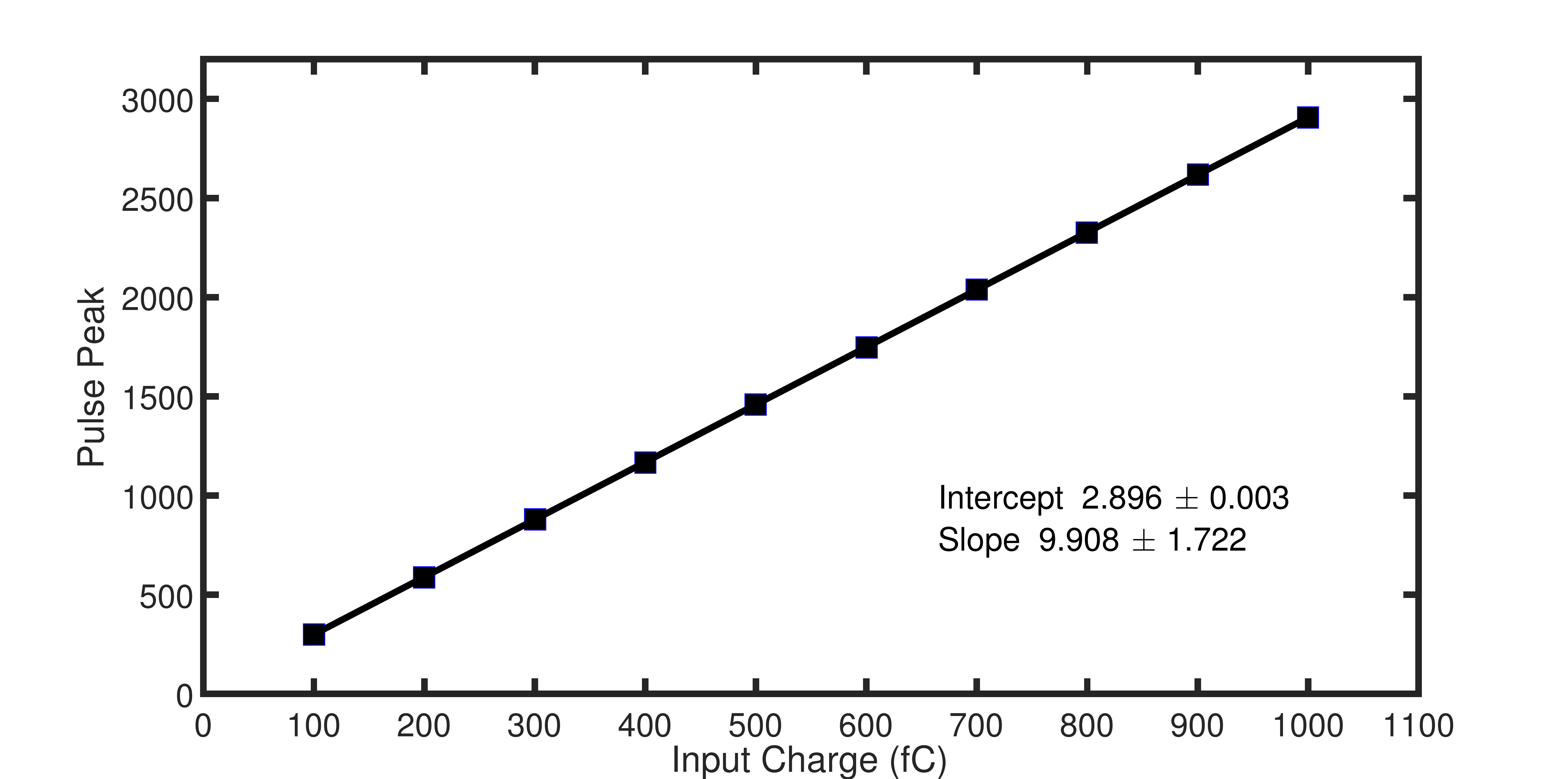}
\caption{Photograph of the typical input-output curve, and the INL is less than 1\%.}
\label{Fig. 8}
\end{figure}

\begin{figure}
\includegraphics[width=3.5in,clip,keepaspectratio]{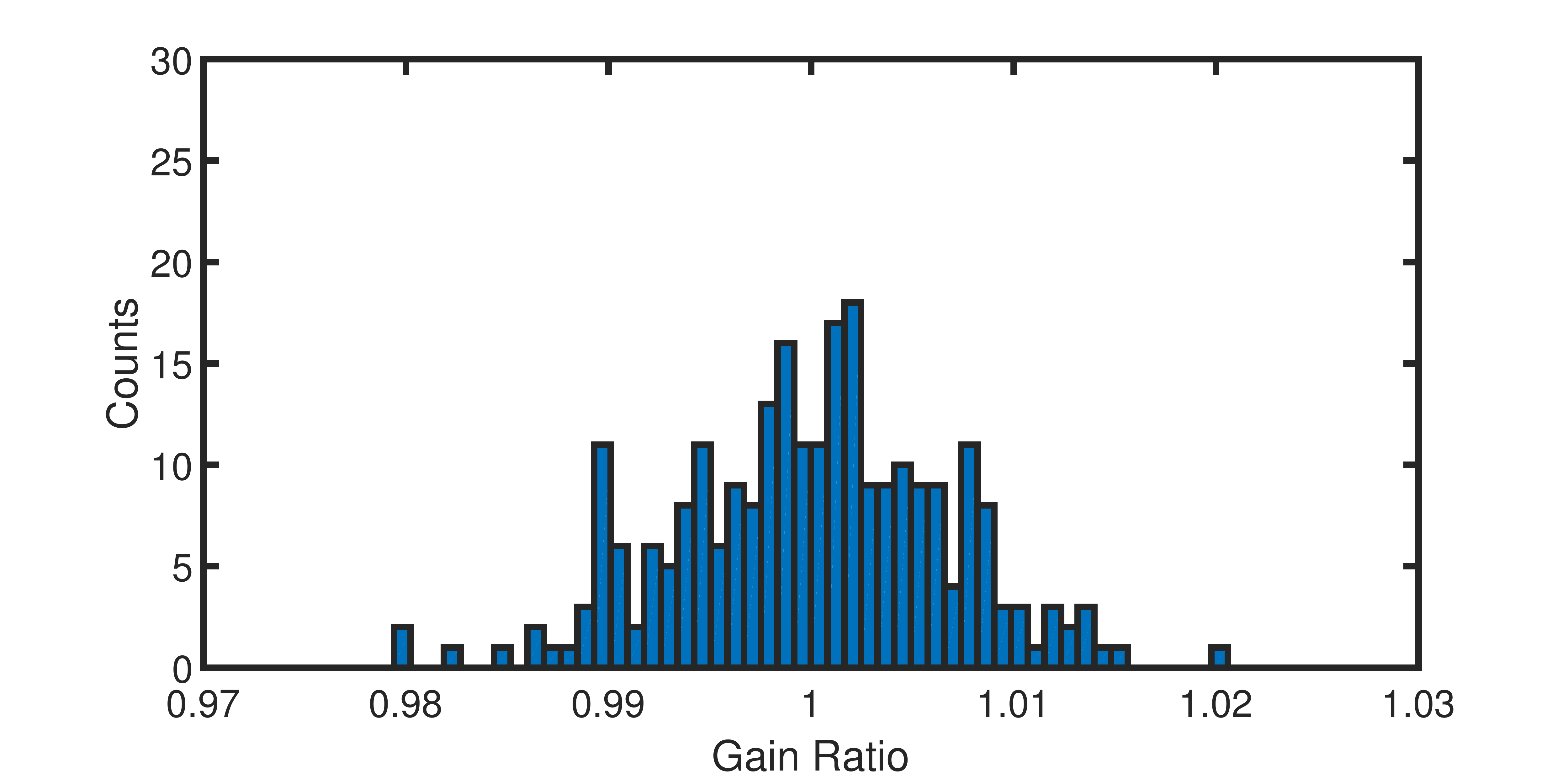}
\caption{Gain distribution of 256 channels, and the non uniformity is around 0.7\%.}
\label{Fig. 9}
\end{figure}

To study the performance of the FEC, a set of tests has been developed. First, the INL and noise of the FEC were measured. The Fig.~\ref{Fig. 7} shows the test of the FEC and the DCM in the laboratory. One way to generate negative charge signals is applying voltage pulse signals provided by the waveform generator AFG3252 to 1 pf capacitors of 5\% precision on the interface board, which is designed to connect the pulse generator with the FEC. Besides, the charge signals can also be generated by the on-board calibration circuit, which produces voltage pulse and sends the signal to the capacitor connected to AGET chips. By adjusting the amplitude of the input voltage pulse, the input-output curve can be plotted, as shown in Fig.~\ref{Fig. 8}, and the INL is less than 1\% with 1 pC range. And the gain non uniformity of 256 channels is around 0.7\%, as shown in Fig.~\ref{Fig. 9}. 


\begin{figure}
\includegraphics[width=3.5in,clip,keepaspectratio]{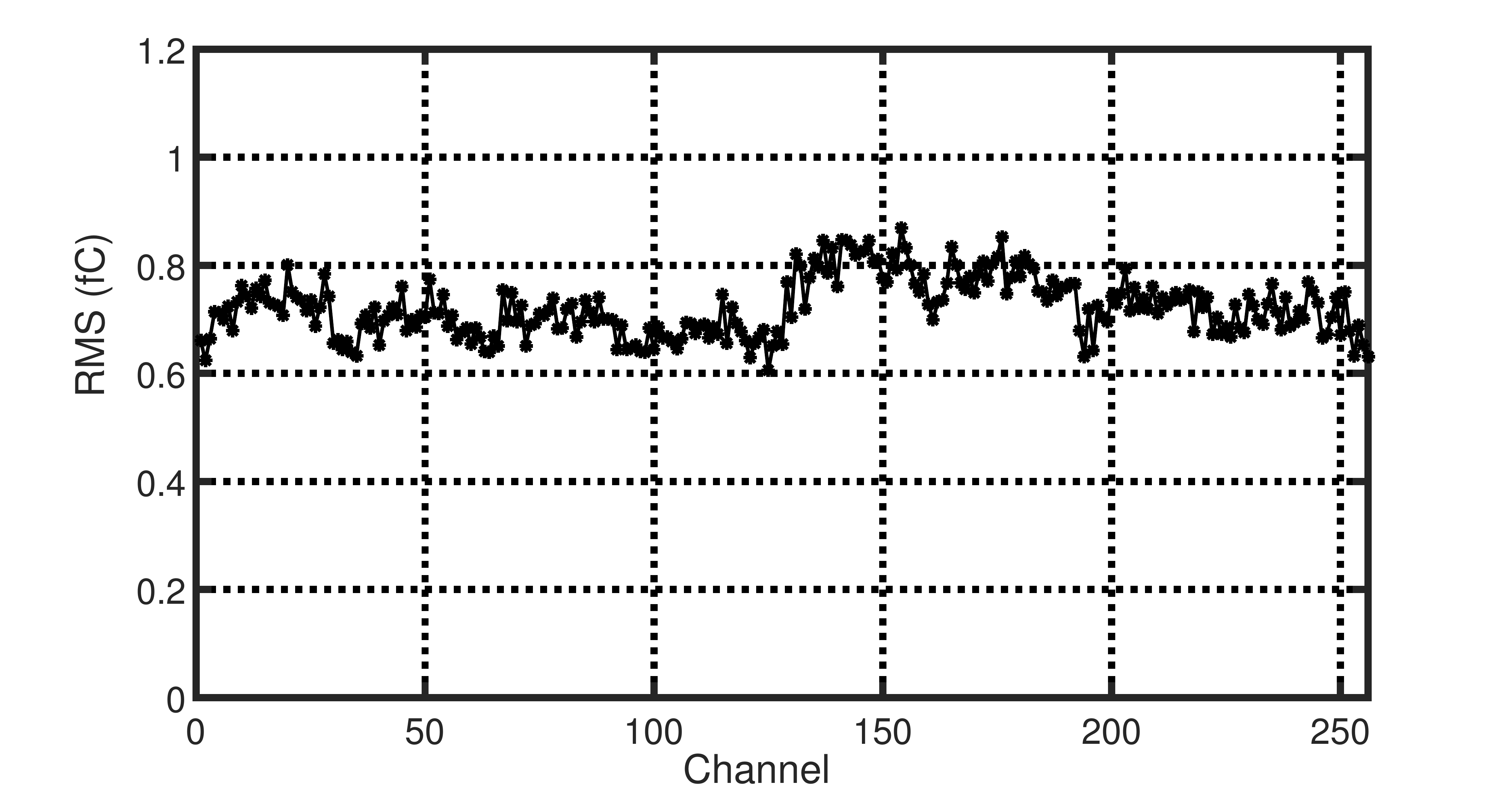}
\caption{Noise (RMS) distribution of 256 channels with 1 $\mu$s peaking time and 1 pC range at room temperature, and the RMS value is less than 0.9 fC.}
\label{Fig. 10}
\end{figure}

Suspending the input channel, the noise distribution of 256 channels on one FEC can be plotted, as shown in Fig.~\ref{Fig. 10}. When setting gain range to 1 pC, the noise (RMS) of 256 channels is less than 2.7 ADC code (0.9 fC) with 1 $\mu$s peaking time. 

\begin{figure}
\includegraphics[width=3.5in,clip,keepaspectratio]{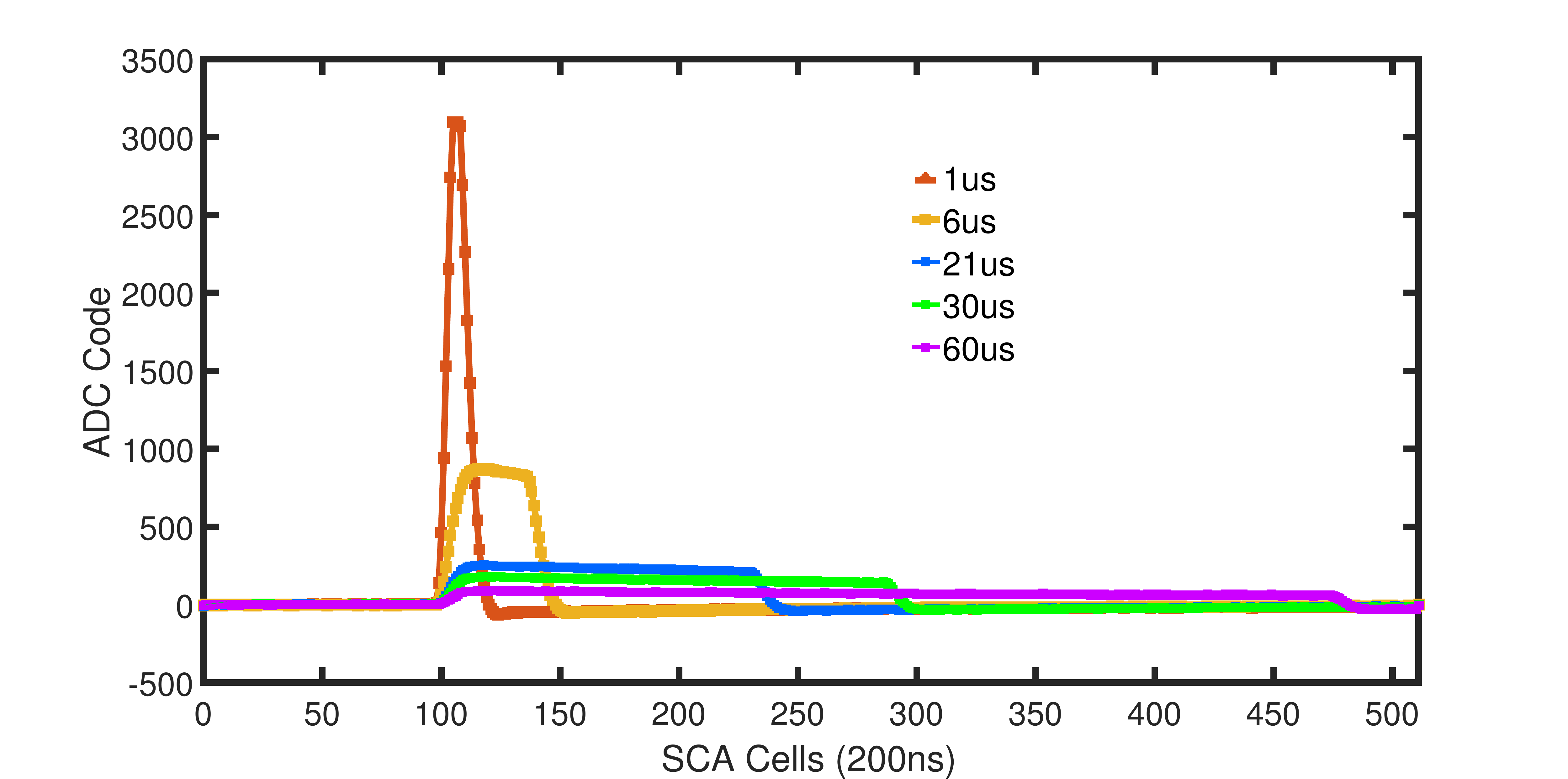}
\caption{The response diagram of different wide pulse on the FEC.}
\label{Fig. 11}
\end{figure}

As the pulse width of anode charge signals can reach 46 $\mu$s, it is crucial to test the response of the FEC to wide pulse
signals. The interface board sends the input pulse to the FEC via pulse generator. The width of the input pulse is adjusted from 1 $\mu$s to 60 
$\mu$s by changing the edge of the pulse on the condition that the charge of the pulse is fixed to 1 pC. The analog signals are processed by the AGET chip and sampled by the ADC, and the test result proves the FEC can handle the wide pulse, as shown in Fig.~\ref{Fig. 11}. 

\begin{figure}
\includegraphics[width=3.6in,clip,keepaspectratio]{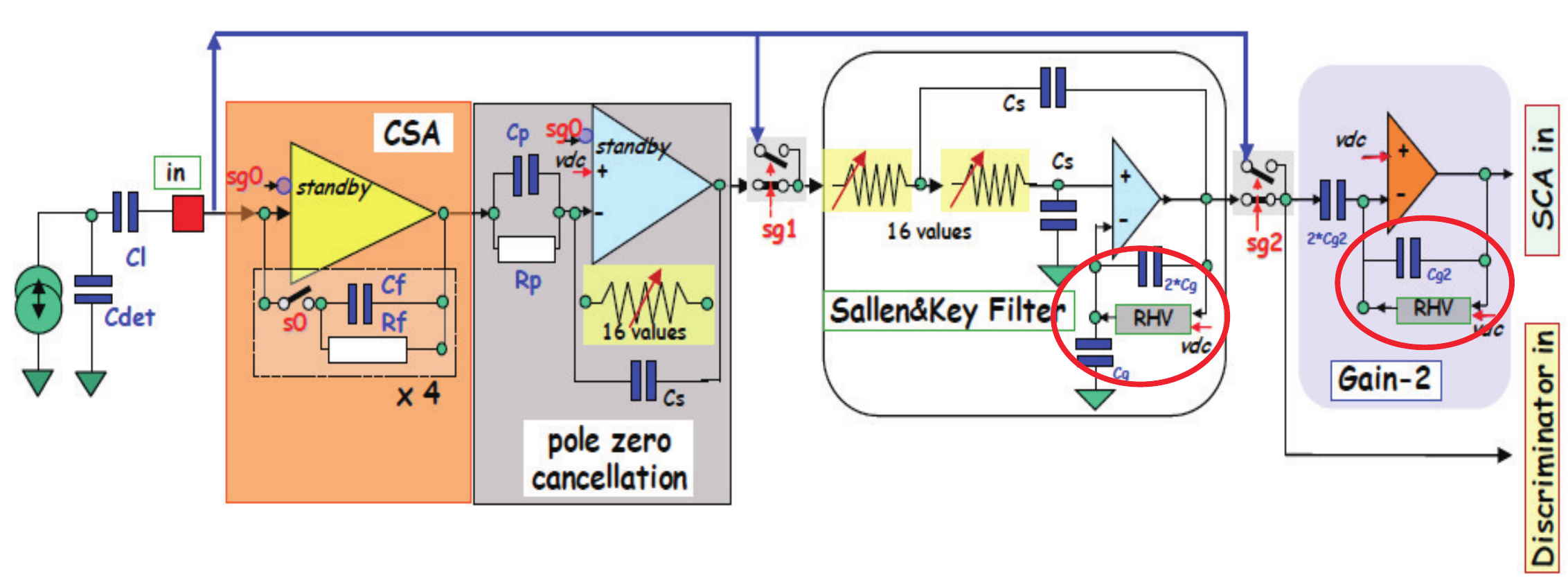}
\caption{Schematic of the front-end part of the analog channel in AGET chip, the figure is from [10].}
\label{Fig. 12}
\end{figure}

\begin{figure}
\includegraphics[width=3.5in,clip,keepaspectratio]{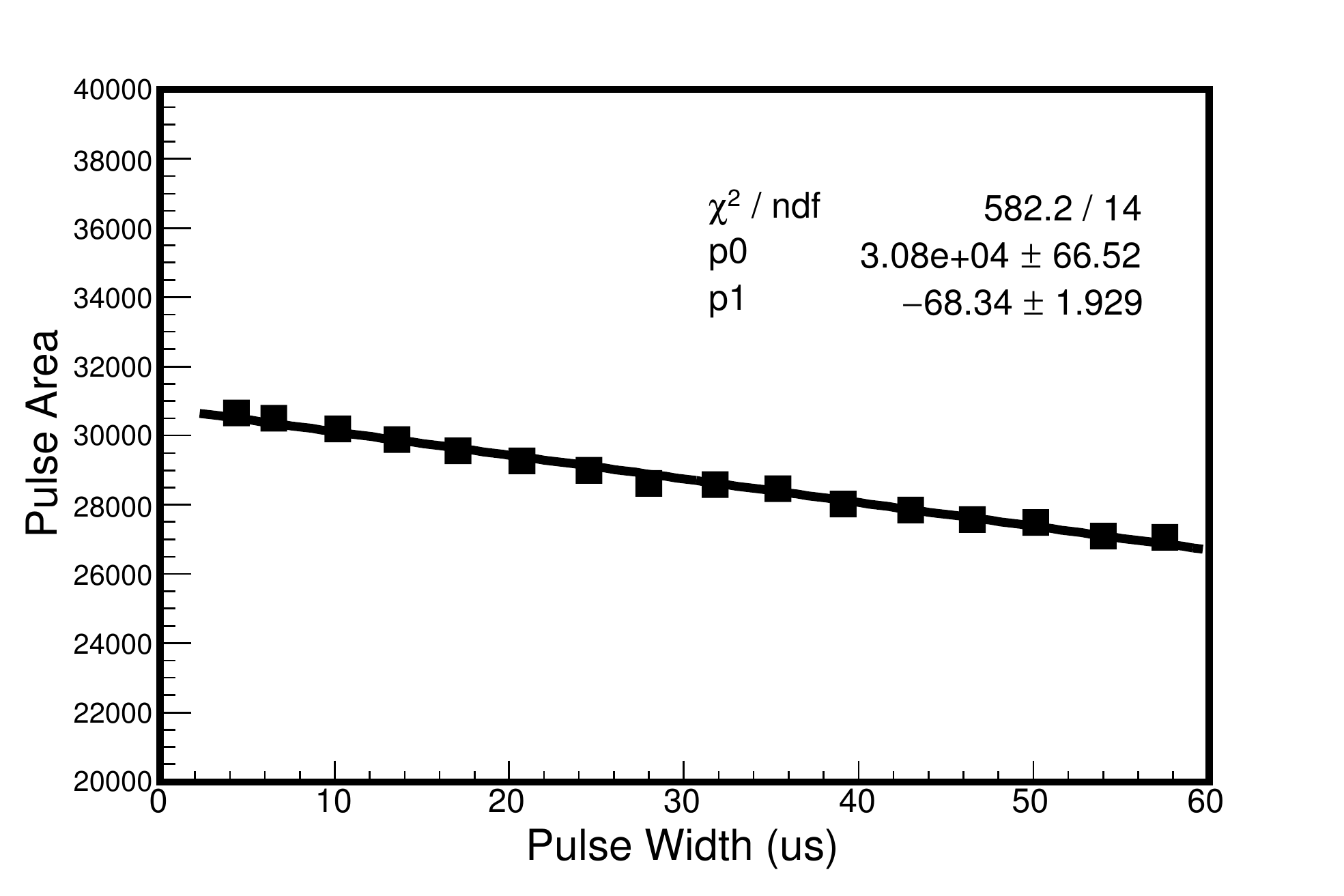}
\caption{The relation diagram of pulse area and pulse width.}
\label{Fig. 13}
\end{figure}

But it is found that the pulse area decreases with the increase of the pulse width, which may arise from the unexpected drop of the flat peak in wide pulse readout. The reason has been found by analyzing the architecture of the analog channel in the AGET chip (Fig.~\ref{Fig. 12}). In order to reduce noise, the capacitors are used instead of resistances in the Sallen\&Key Filter and Gain-2 Amplifier to realize the negative feedback. Thus the use of the nonlinear components can cause the decrease of the magnification at low frequency. Fortunately, the attenuation of the pulse area follows the linear fit (Fig.~\ref{Fig. 13}) so that it can be calibrated and corrected in the experiment.

Finally, the reliability of optical link between the FEC and the DCM was also tested. The result of bit error test shows the bit error rate is acceptable, as shown in Table~\ref{Table I}. More bit error tests of long time will be carried out in the future. 

\begin{table}[htbp]
\caption{The result of bit error test between FEC and the DCM.}
\scriptsize
\centering
\label{Table I}
\begin{tabular*}{3.5in}{p{20pt}p{33pt}p{25pt}p{25pt}p{25pt}p{55pt}}
\hline
\hline
\\
\begin{center}
Receiver
\end{center}&
\begin{center}
Transmission
\end{center}&
\begin{center}
Test Time
\end{center}&
\begin{center}
Test Bit Num
\end{center}&
\begin{center}
Bit Err Num
\end{center}&
\begin{center}
Bit Error Rate (90\% Confidence) 
\end{center}\\

\\
\hline\\
FEC&DCM&24h&70$\times10^{13}$&0&$<$3.2$\times10^{-14}$ \\
\\
DCM&FEC&24h&70$\times10^{13}$&0&$<$3.2$\times10^{-14}$  \\
\\
\hline
\hline
\end{tabular*}
\end{table}

\section{Joint Test with the Prototype TPC}
\begin{figure}
\includegraphics[width=3.5in,clip,keepaspectratio]{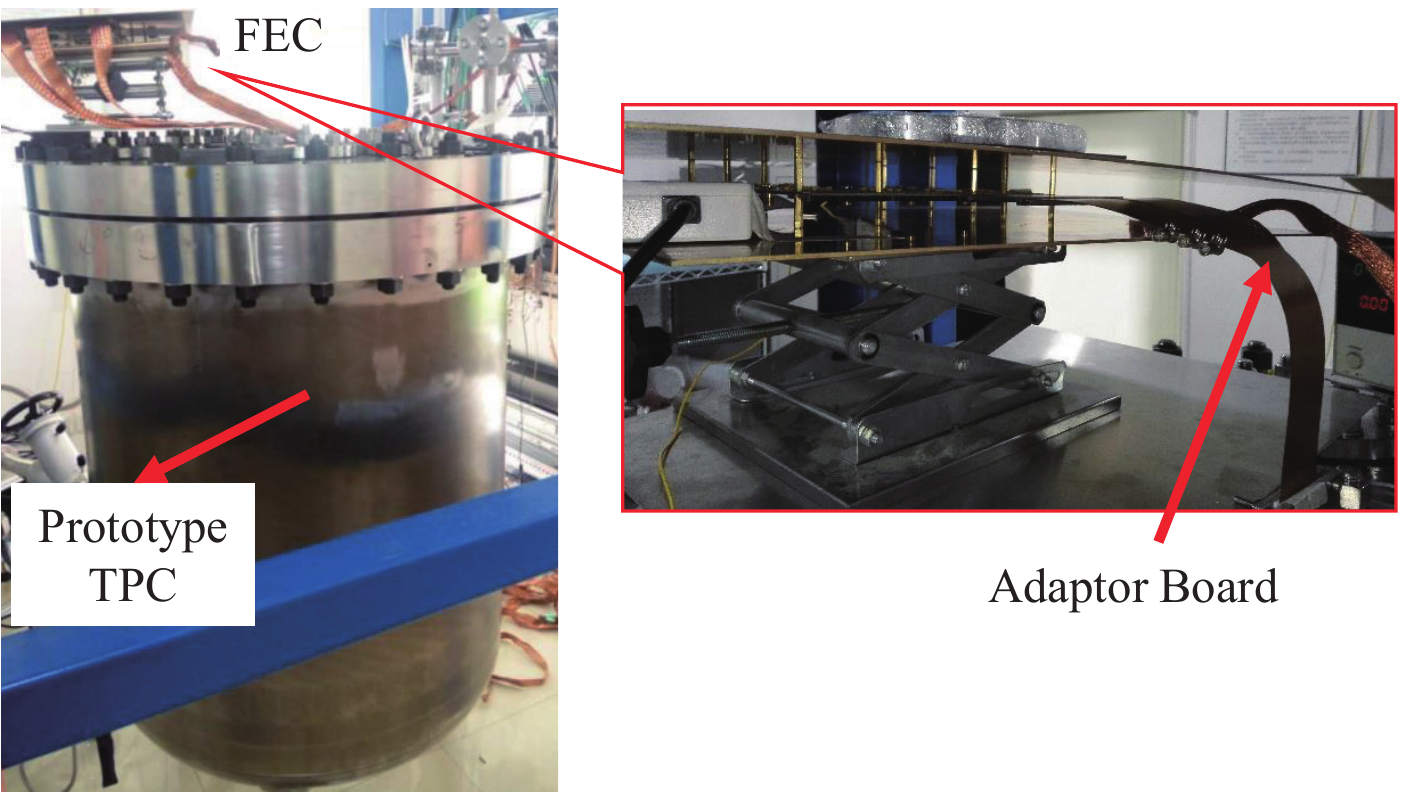}
\caption{A photograph of one FEC developed in joint test with the prototype TPC.}
\label{Fig. 14}
\end{figure}

\begin{figure}
\includegraphics[width=3.5in,clip,keepaspectratio]{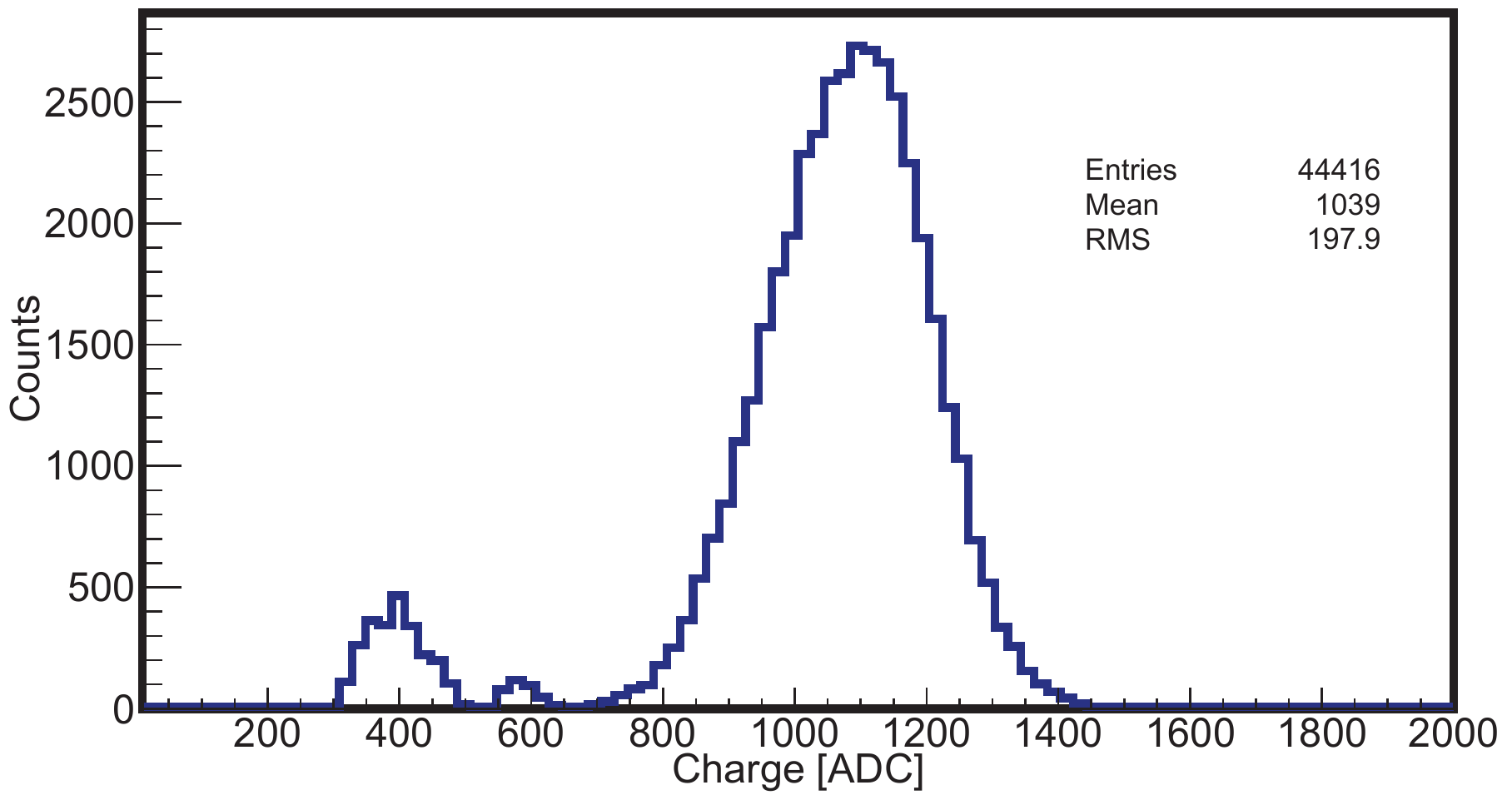}
\caption{Energy spectrum of $^{56}$Fe reconstructed in the joint-test with prototype TPC detector.}
\label{Fig. 15}
\end{figure}

To evaluate the performance of the FEC for the experiment, one FEC was mounted on the prototype TPC detector to readout 128 channels through an adaptor board at first, as shown in Fig.~\ref{Fig. 14}. The applied voltage on the TPC cathode is -12 kV and that on the Micromesh is -340 V. Thus the gain of the Micromegas detector is around 1500 during the joint-test. The mixture gas filled in the detector is argon (Ar, 95\%) and isobutane (C4H10, 5\%) at 1 bar. The radioactive source used here is $^{56}$Fe to test the performance of the detector. During the test, the sampling frequency of the AGET chip is set to 5 MHz and peaking time is 1 $\mu$s with 120 fC input range. Its preliminary energy spectrum is shown in Fig.~\ref{Fig. 15}, and the energy resolution is 27\% to 28\% @ 5.9 KeV. 

\begin{figure}
\includegraphics[width=3.5in,clip,keepaspectratio]{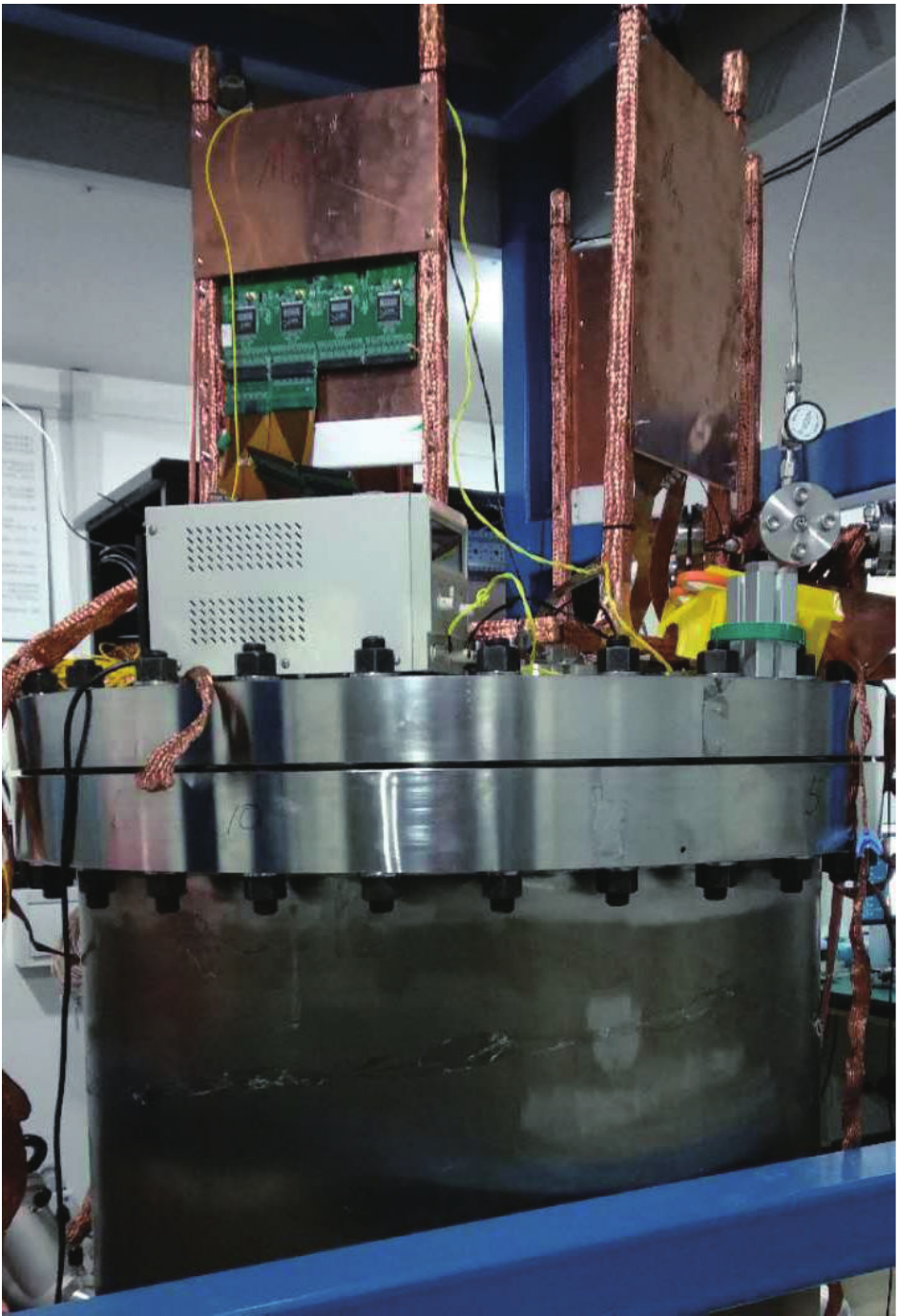}
\caption{A photograph of four FECs developed in the joint-test with the prototype TPC.}
\label{Fig. 16}
\end{figure}

The prototype TPC in SJTU is equipped with 7 Microbulk Micromegas modules so far, but currently there are only 6 Micromegas modules in good operation. So four FECs were placed on the outer surface of the TPC end-cap to readout the 768 channels through 6 adaptor boards. The testing of the system just started. The applied voltage on the TPC cathode is -40 kV and that on the Micromesh is -550 V. Thus the gain of the Micromegas detector is around 300 during the joint-test. The mixture gas filled in the detector is argon (Ar, 98.5\%) and isobutane (C4H10, 1.5\%) at 5 bar. The radioactive source used here is $^{137}$Cs to test the performance of the detector. The location of $^{137}$Cs is in the middle of Micromegas 6 detector (M6), also the center of all Micromegas detectors. The joint-test of FECs with prototype TPC has been developed in SJTU (Fig.~\ref{Fig. 16}). 

\begin{figure}
\includegraphics[width=3.5in,clip,keepaspectratio]{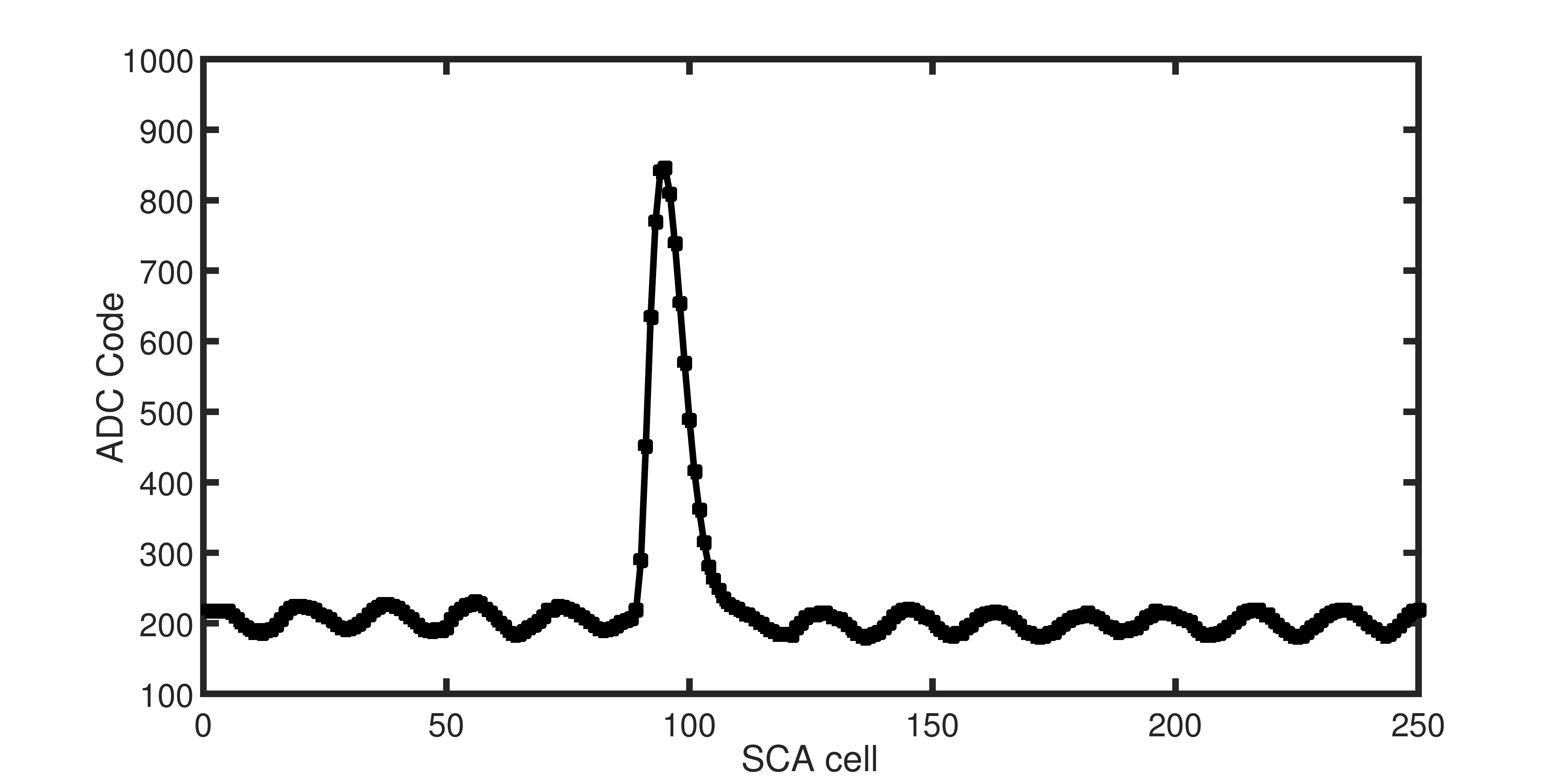}
\caption{A typical waveform of the detector signal in one readout node.}
\label{Fig. 17}
\end{figure}

\begin{figure}
\includegraphics[width=3.5in,clip,keepaspectratio]{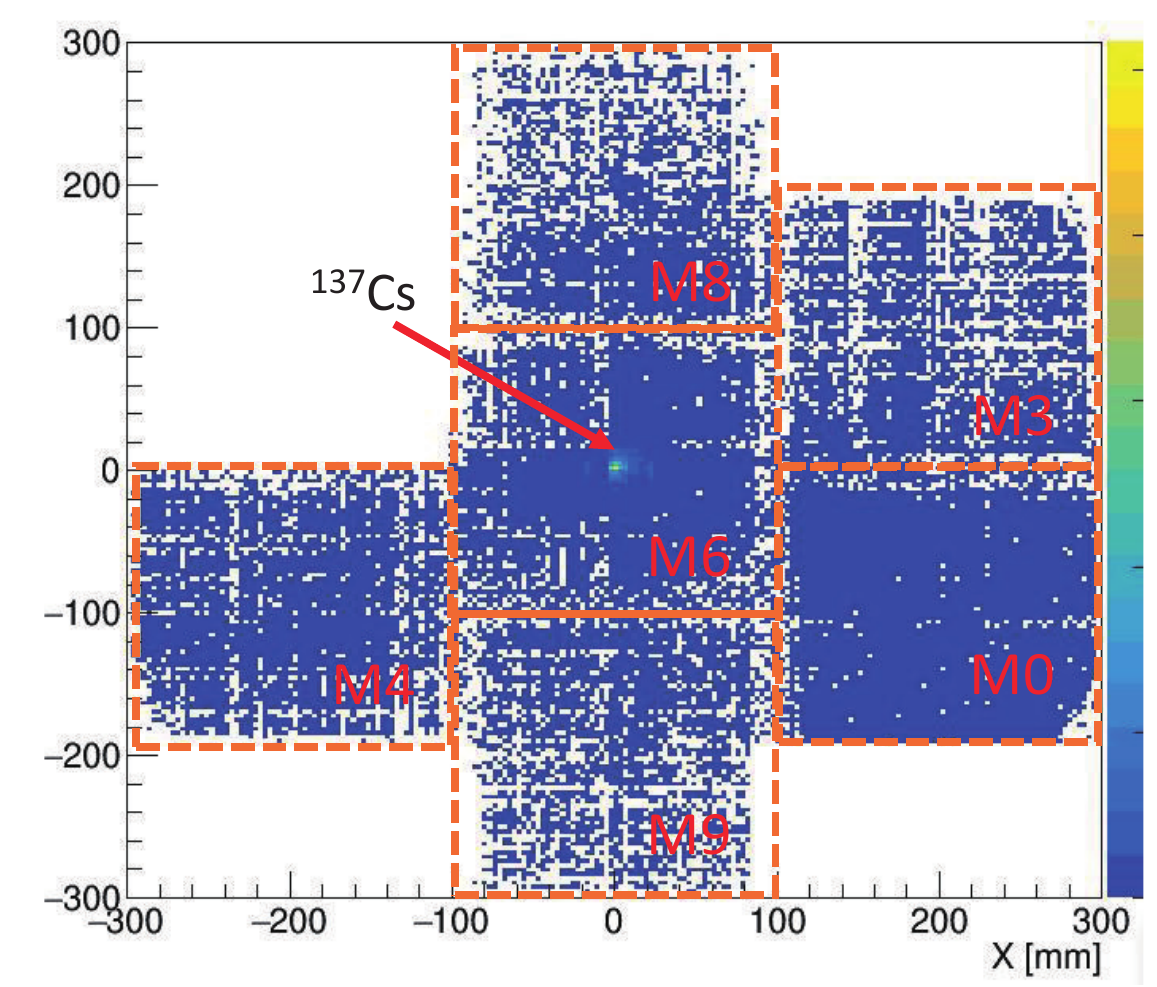}
\caption{Hit map of 6 Micromegas in the join-test under the radiation source $^{137}$Cs.}
\label{Fig. 18}
\end{figure}

During the experiment, the sampling frequency of the AGET chip is set to 5 MHz and peaking time is 1 $\mu$s with 1 pC input range. Fig.~\ref{Fig. 17} shows a typical plot of the detector signal readout by the anode. The peak and area of pulse can be used as its energy information. After data analysis, the hit map (Fig.~\ref{Fig. 18}) shows the position of six Micromegas detectors and a clear vision of radiation source can be seen. The result shows the performance of FEC meets the requirements of the prototype TPC.

\section{Conclusion}
The key design and performance of the front-end electronics for PandaX-III prototype TPC are presented in this paper. And the performance of front-end card meets the requirements of PandaX-III experiment. During the joint-test with PandaX-III prototype TPC, the gain range of all FECs can be set to 1 pC, and the sampling rate can be set to 5 MHz. The FECs with 1024 readout channels function well and the hit map of 6 Micromegas has been reconstructed.


%



\section*{Acknowledgment}
We would like to thank all the colleagues from PandaX-III collaboration for their helpful support. Shaobo Wang from Shanhai Jiaotong University helped us with the joint test and provided some results of the test. And we also want to thank CALVET Denis from the Saclay laboratory for his helpful suggestions on application and test of the AGET chip.

\ifCLASSOPTIONcaptionsoff
  \newpage
\fi

\end{document}